\documentclass[11pt, a4paper]{article}

\usepackage{natbib}
\usepackage{bm}
\usepackage[left = 3cm, right = 3cm, top = 2.5cm, bottom = 2.5cm]{geometry}
\usepackage{amsmath}
\usepackage{tabularx}
\usepackage{amssymb}
\usepackage{amsthm}
\usepackage{pdfsync}
\usepackage{booktabs}
\usepackage{multirow}
\usepackage{graphicx}
\usepackage{mathdots}
\usepackage{url}
\usepackage{hyperref}

\title{Bias in parametric estimation: \\ reduction and useful
  side-effects}
\author{Ioannis Kosmidis \\ Department of Statistical Science, University College London}

\begin{document}

\def\var{\mathop{\rm var}}
\def\cov{\mathop{\rm cov}}
\def\cum{\mathop{\rm cum}_3}
\def\expect{E}
\def\log{\mathop{\rm log}}
\def\trace{\mathop{\rm tr}}
\def\asvec{\mathop{\rm vec}}
\def\bigo{O}   
\def\diag{\mathop{\rm diag}}
\def\ahalf{\frac{1}{2}}

\maketitle

\begin{abstract}
  The bias of an estimator is defined as the difference of its
  expected value from the parameter to be estimated, where the
  expectation is with respect to the model. Loosely speaking, small
  bias reflects the desire that if an experiment is repeated
  indefinitely then the average of all the resultant estimates will be
  close to the parameter value that is estimated. The current paper is
  a review of the still-expanding repository of methods that have been
  developed to reduce bias in the estimation of parametric models. The
  review provides a unifying framework where all those methods are
  seen as attempts to approximate the solution of a simple estimating
  equation. Of particular focus is the maximum likelihood estimator,
  which despite being asymptotically unbiased under the usual
  regularity conditions, has finite-sample bias that can result in
  significant loss of performance of standard inferential procedures.
  An informal comparison of the methods is made revealing some useful
  practical side-effects in the estimation of popular models in
  practice including: i) shrinkage of the estimators in binomial and
  multinomial regression models that guarantees finiteness even in
  cases of data separation where the maximum likelihood estimator is
  infinite, and ii) inferential benefits for models that require the
  estimation of dispersion or precision parameters. 
\end{abstract}

\section{Impact of bias in estimation}
By its definition, bias necessarily depends on how the model is
written in terms of its parameters and this dependence makes it not a
strong statistical principle in terms of evaluating the performance of
estimators; for example, unbiasedness of the familiar sample variance
$S^2$ as an estimator of $\sigma^2$ does not deliver an unbiased
estimator of $\sigma$ itself. Despite this fact, an extensive amount
of literature has focused on unbiased estimators (estimators with zero
bias) as the basis of refined statistical procedures (for example,
finding minimum variance unbiased estimators). In such work
unbiasedness plays the dual role of a condition i) that allows the restriction of the class of possible estimators in order to obtain something useful
(like minimum variance amongst unbiased estimators), and ii) that
ensures that estimation is performed in an impartial way, ruling out
estimators that would favour one or more parameter values at the cost
of neglecting other possible values. \citet[Chapter~2]{lehmann:98} is
a thorough review of statistical methods that are optimal once
attention is restricted to unbiased estimators.

Another stream of literature has focused in reducing the bias of
estimators, as a means to alleviating the sometimes considerable
problems that bias can cause in inference. This literature, despite
dating back to the early years of statistical science, is resurfacing
as increasingly relevant as the complexity of models used in practice
increases and pushes traditional estimation methods to their
theoretical limits.

The current review focuses on the latter literature,
explaining the link between the available methods for bias reduction
and their relative merits and disadvantages through the analysis of
real data sets.

The following case study demonstrates the direct consequences that the
bias in the estimation of a single nuisance (or incidental) parameter
can have in inference, even if all parameters of interest are
estimated with negligible bias.

\subsection{Gasoline yield data}
\label{gasoline}
To demonstrate how bias can in some cases severely affect estimation
and inference we follow the gasoline yield data example in
\citet{kosmidis:11} and \citet{grun:12}. The gasoline yield data
\citep{prater:56} consists of $n = 32$ observations on the proportion
of crude oil converted to gasoline after distillation and
fractionation on $10$ distinct experimental settings for the triplet
i) temperature in degrees Fahrenheit at which 10\% of crude oil has
vaporized, ii) crude oil gravity, and iii) vapor pressure of crude
oil. The temperature at which all gasoline has vaporized is also
recorded in degrees Fahrenheit for each one of the $32$ observations.

The task is to fit a statistical model that links the proportion of
crude oil converted to gasoline with the experimental settings and the
temperature at which all gasoline has vaporized. For this we assume
that the observed proportions of crude oil converted to gasoline $y_1,
y_2, \ldots, y_{n}$ are realizations of independent Beta distributed
random variables $Y_1, \ldots, Y_{n}$, where $\mu_i = E(Y_i)$ and
$\var(Y_i) = \mu_i(1 - \mu_i)/(1+\phi)$. Hence, in this
parameterization, $\phi$ is a precision parameter $(i = 1,\ldots,
n)$. Then, the mean $\mu_i$ of the $i$th response can be linked to a linear
combination of covariates and regression parameters via the logistic
link function as
  \begin{equation}
    \label{gasolineModel}
    \log{\frac{\mu_i}{1-\mu_i}} = \alpha + \sum_{t =
      1}^{9}\gamma_t s_{it} + \delta t_i \quad (i = 1, \ldots, n) \, .
  \end{equation}
  In the above expression, $s_{i1}, \ldots, s_{i9}$ are the values of 9
  dummy covariates which represent the 10 distinct experimental
  settings in the data set and $t_{i}$ is the temperature in degrees
  Fahrenheit at which all gasoline has vaporized for the $i$th
  observation $(i =1, \ldots, n)$.

  The parameters $\theta = (\alpha, \gamma_1, \ldots, \gamma_{9},
  \delta, \phi)$ are estimated using maximum likelihood and the
  estimated standard errors for the estimates are calculated using the
  square roots of the diagonal elements of the inverse of the Fisher
  information matrix for model (\ref{gasolineModel}). The parameter
  $\phi$ is considered here to be a nuisance (or incidental) parameter
  which is only estimated to complete the specification of the Beta
  regression model.

\begin{table}[t]
\caption{Maximum likelihood estimates for the parameters of
  model~(\ref{gasolineModel}) with the corresponding estimated
  standard errors and the Wald-type $95\%$ confidence intervals
  (``estimate'' $\pm$ 1.96 ``estimated standard
  error'').}
\begin{center}
\begin{tabular}{ccccc}
  \toprule
  Parameter & Estimate & Estimated standard error &
  \multicolumn{2}{c}{95\% confidence interval} \\
  \midrule
  $\alpha$  &   -6.160 &    0.182 &   -6.517 &   -5.802 \\
  $\gamma_1$  &    1.728 &    0.101 &    1.529 &    1.926 \\
  $\gamma_2$  &    1.323 &    0.118 &    1.092 &    1.554 \\
  $\gamma_3$  &    1.572 &    0.116 &    1.345 &    1.800 \\
  $\gamma_4$  &    1.060 &    0.102 &    0.859 &    1.260 \\
  $\gamma_5$  &    1.134 &    0.104 &    0.931 &    1.337 \\
  $\gamma_6$  &    1.040 &    0.106 &    0.832 &    1.248 \\
  $\gamma_7$  &    0.544 &    0.109 &    0.330 &    0.758 \\
  $\gamma_8$  &    0.496 &    0.109 &    0.282 &    0.709 \\
  $\gamma_9$  &    0.386 &    0.119 &    0.153 &    0.618 \\
  $\delta$  &    0.011 &    0.000 &    0.010 &    0.012 \\
  $\phi$  &  440.278 &  110.026 &  224.632 &  655.925 \\
\bottomrule
\end{tabular}
\end{center}
\label{gasolineML}
\end{table}

Table~\ref{gasolineML} shows the parameter estimates with the
corresponding estimated standard errors and the $95\%$ Wald-type
confidence intervals. One immediate observation from the table of
coefficients is the very large estimate for the precision parameter
$\phi$. If this is merely the effect of upward bias then this bias
will result in underestimation of the standard errors because for such
a model the entries of the Fisher information matrix corresponding to
the regression parameters $\alpha, \gamma_1, \ldots, \gamma_{9},
\delta$ are quantities of the form ``$\phi$ times a function of
$\theta$'' \citep[see,][for expressions on the Fisher
information]{kosmidis:11, grun:12}. Hence, if the estimation of $\phi$
is prone to upward bias, then this can lead to confidence intervals
that are shorter than expected at any specified nominal level and/or
anti-conservative hypothesis testing procedures, which in turn result
in spuriously strong conclusions.

To check whether this is indeed the case a small simulation study has
been designed where $50000$ samples are simulated from the maximum
likelihood fit shown in Table~\ref{gasolineML}. Maximum likelihood is
used to fit model~(\ref{gasolineModel}) on each simulated sample
and the bias of the maximum likelihood estimator is estimated using
the resultant parameter estimates. The estimated bias for $\alpha$ is
$0.010$ while the estimated biases for $\gamma_1, \ldots, \gamma_{9},
\delta$ are all less than $0.005$ in absolute value, providing
indications that bias on the regression parameters is of no
consequence. Nevertheless, the estimated bias for $\phi$ is $299.779$
which indicates a strong upward bias in the estimation of $\phi$. To
check how the upward bias in the precision parameter can affect the
usual Wald-type inferences, we estimate the coverage probability (the
probability that the confidence intervals contains the true parameter
value) of the individual Wald-Type confidence intervals at levels
$90\%$, $95\%$ and $99\%$. Table~\ref{gasolineCover} shows the
results. It is clear that the Wald-type confidence intervals
systematically undercover the true parameter value across parameters.

\begin{table}[t]
\caption{Estimated coverage of Wald-type
  confidence intervals at nominal level $90\%$, $95\%$ and
  $99\%$. Estimated standard errors are calculated using the Fisher
  information at the maximum likelihood estimates.}
\begin{center}
\begin{tabular}{cccc}
  \toprule
  Parameter & \multicolumn{3}{c}{Level} \\
	&  90\% & 95\% & 99\% \\
  \midrule
$\alpha$   & 80.2\% &  87.2\% &  94.9\% \\
$\gamma_1$ & 80.3\% &  87.3\% &  95.2\% \\
$\gamma_2$ & 80.2\% &  87.1\% &  95.1\% \\
$\gamma_3$ & 80.2\% &  87.1\% &  94.8\% \\
$\gamma_4$ & 80.2\% &  87.5\% &  95.2\% \\
$\gamma_5$ & 80.5\% &  87.5\% &  95.2\% \\
$\gamma_6$ & 80.4\% &  87.4\% &  95.1\% \\
$\gamma_7$ & 80.6\% &  87.4\% &  95.1\% \\
$\gamma_8$ & 80.2\% &  87.3\% &  95.1\% \\
$\gamma_9$ & 80.5\% &  87.3\% &  95.0\% \\
$\delta$   & 79.9\% &  87.1\% &  94.9\% \\
\bottomrule
\end{tabular}
\end{center}
\label{gasolineCover}
\end{table}


Such behaviour is observed even when the precision parameter is linked
to covariates through a link function, like logarithm \citep[see, for
example][]{grun:12}. More generally, similar consequences of bias in
inference are present in all exponential family models that involve the
estimation of dispersion (or precision) parameters.

\section{Consistency, bias and variance}
Suppose that interest is in the estimation of a $p$-vector of
parameters $\theta$, from data $y^{(n)}$ assumed to be realizations of
a random quantity $Y^{(n)}$ distributed according to a parametric
distribution $M_\theta$, $\theta = (\theta_1, \ldots, \theta_p)^T \in
\Theta \subset \Re^p$. The superscript $n$ here is used as an
indication of the information in the data and is usually the sample
size in the sense that the realization of $Y^{(n)}$ is $y^{(n)} =
(y_1, \ldots, y_n)^T$. An estimator of $\theta$ is a function
$\hat\theta \equiv t(Y^{(n)})$ and in the presence of data the
estimate would be $t(y^{(n)})$.


An estimator $\hat\theta = t(Y^{(n)})$ is consistent if it converges
in probability to the unknown parameter $\theta$ as $n \rightarrow
\infty$. Consistency is usually an essential requirement for a good
estimator because given that the family of distributions $M_\theta$ is
large enough, it ensures that as $n$ increases the distribution of
$\hat\theta$ becomes concentrated around the parameter $\theta$,
essentially providing a practical reassurance that for very large $n$
the estimator recovers $\theta$.

The bias of an estimator is defined as
\[
B(\theta) = E_\theta(\hat\theta - \theta) \, .
\]
Loosely speaking, small bias reflects the desire that if an experiment
that results in data $y^{(n)}$ is repeated indefinitely, then the
long-run average of all the resultant estimates will not be far from
$\theta$. Small bias is a much weaker and hence less useful
requirement than consistency. Indeed, one may get an inconsistent
estimator with zero bias or a consistent estimator that is biased.
For example, if $Y^{(n)} = (Y_1, \ldots, Y_n)^T$, with $Y_1, \ldots,
Y_n$ mutually independent random variables with $Y_i \sim N(\mu,
\sigma^2)$ then $t(Y^{(n)}) = Y_1$ is an unbiased but inconsistent
estimator of $\mu$.  On the other hand, $t(Y^{(n)}) =
\sum_{i=1}^n{Y_i} + 1/n$ is a consistent estimator for $\mu$ but has
bias $B(t(Y^{(n)})) = 1/n$.  So, bias becomes relevant only if it is
accompanied by guarantees of consistency, or more generally when the
variability of $\hat\theta$ around $\theta$ is small \citep[see,][\S
8.1 for a discussion along this lines]{cox:74}.


The bias function also appears directly in the expression for the
lowest attainable variance of an estimator. The Cram\'{e}r-Rao inequality
states that the variance of any estimator $\hat\theta^{(n)}$ satisfies
\begin{equation}
\label{CRbound}
\var(\hat\theta^{(n)}) \succeq \left\{1_p + \nabla_\theta
  B(\theta)\right\}^T\left\{F(\theta)\right\}^{-1}\left\{1_p
  + \nabla_\theta B(\theta)\right\}\, ,
\end{equation}
where $1_p$ is the $p \times p$ identity matrix and the inequality $A
\succeq C$ means that $A - C$ is a positive semi-definite
matrix. The matrix $F(\theta)$ is the Fisher (or expected) information
matrix which is defined as $F(\theta) = E_\theta\left\{S(\theta)
  S(\theta)^T\right\}$, where $S(\theta) = \nabla_\theta l(\theta)$
and $l(\theta)$ is the log-likelihood function for $\theta$. The
Cram\'{e}r-Rao inequality shows what is the ``lowest'' attainable variance for
an estimator in terms of the derivatives of its bias and the Fisher
information.

\subsection{Maximum likelihood estimation}
Denote $f(y; \theta)$ the joint density or probability mass function
implied by the family of distributions $M_\theta$. The maximum
likelihood estimator $\hat\theta$ is the value of $\theta$ which
maximizes the log-likelihood function $l(\theta; y^{(n)}) = \log
f(y^{(n)}; \theta)$. Given that the log-likelihood function is
sufficiently smooth on $\theta$, $\hat\theta$ can be obtained as the
solution of the score equations
\[
S(\theta) = \nabla_\theta l(\theta) = 0 \, ,
\]
provided that the observed information matrix $I(\theta) = -
\nabla_\theta\nabla_\theta^T l(\theta)$ is positive definite when
evaluated at $\hat\theta$.  An appealing property of the maximum
likelihood estimator is its invariance under one-to-one
reparameterizations of the model. If $\theta' = g(\theta)$ for some
one-to-one function $g : \Re^p \to \Re^p$, then the maximum likelihood
estimator of $\theta'$ is simply $g(\hat\theta)$. This result states
that when obtaining the maximum likelihood estimator of $\theta$, we
automatically obtain the maximum likelihood estimator of $g(\theta)$
for any function $g$ that is one-to-one, simply by calculating
$g(\hat\theta)$ without the need of maximizing the likelihood on
$g(\theta)$.

It can also be shown that the maximum likelihood estimator
$\hat\theta$ has certain optimality properties if the ``usual
regularity conditions'' are satisfied. Informally, the usual
regularity conditions imply, amongst others, that i) $M_\theta$ is
identifiable (that is $M_\theta \ne M_{\theta'}$, for any pair
$(\theta, \theta')$ such that $\theta \ne \theta'$, apart from sets of
probability zero), ii) $p$ is finite, iii) that the parameter space
$\Theta$ does not depend on the sample space (which implies that $p$
does not depend on $n$ and iv) that there exists a sufficient number
of log-likelihood derivatives and expectations of those under
$M_\theta$. A more technical account of those conditions can be found
in \citet[\S\S 7.1,7.2]{mccullagh:87}, or equivalently in \citet[\S\S
9.1]{cox:74}.

If these conditions are satisfied, then $\hat\theta$ is consistent and
has bias of asymptotic order $O(n^{-1})$, which means that its bias vanishes as
  $n \to \infty$. Moreover, the maximum likelihood estimator has the
  property that as $n \to \infty$ its distribution converges to a
  multivariate Normal distribution with expectation $\theta$ and
  variance covariance matrix $\left\{F(\theta)\right\}^{-1}$. Hence,
  the variance of the asymptotic distribution of the maximum
  likelihood estimator is exactly the Cram\'{e}r-Rao lower bound
  $\left\{F(\theta)\right\}^{-1}$ given in (\ref{CRbound}).

\subsection{Reducing bias}
All the above shows that under the usual regularity conditions as $n
\to \infty$, the maximum likelihood estimator $\hat\theta$ has optimal
properties, a fact that makes it a default choice in
applications. However, for finite $n$ these properties may
deteriorate, in some cases causing severe problems in inference. Such an
effect has been seen in the gasoline yield data case study where the
bias of $\hat\theta$ affects the
performance of tests and the construction of confidence intervals
based on the asymptotic Normality of $\hat\theta$.


Before reviewing the basic methods for reducing bias, it is necessary
to emphasize again that bias necessarily depends on the
parameterization of the model; if the bias of any estimator
$\hat\theta$ is reduced resulting to a less biased estimator
$\tilde\theta$, then it is not necessary that the same will happen for
the estimator $g(\tilde\theta)$. In fact, the bias of the
$g(\tilde\theta)$ as an estimator of $g(\theta)$ may be considerably
inflated. Hence, correction of the bias of the maximum likelihood
estimator comes at the cost of destroying its invariance properties
under reparameterization. Therefore, all the methods for bias
reduction that are described in the current review should be seen with
scepticism if invariance is a necessary requirement for the
analysis. On the other hand if the parameterization is fixed by the
problem or practitioner, one can do much better in terms of estimation
and inference by reducing the bias. Furthermore, as it will be seen
later, for some models reduction of bias produces useful side-effects
which in many cases have motivated its routine use in applications. A
thorough discussion on considerations on bias and variance and
examples of exactly unbiased estimators that are useless or irrelevant
can be found at \citet[][\S 8.2]{cox:74} and \citet[\S
1.1]{lehmann:98}.

\section{Bias reduction - A simple recipe with many different implementations}
For a general --- not necessarily the maximum likelihood --- estimator
$\hat\theta$ taking values in $\Theta \subset \Re^p$, consider the
solution of the equation
\begin{equation}
\label{estEq}
\hat\theta - \tilde\theta = B(\theta) \, ,
\end{equation}
with respect to a new estimator $\tilde\theta$. Equation~(\ref{estEq})
is a moment-matching equation which links the properties of the
estimation method to the properties of $M_\theta$ through $\hat\theta$
and $B(\theta)$, respectively. If both the function $B(\theta)$ and
$\theta$ were known then it is straightforward to show that
$\tilde\theta = \hat\theta - B(\theta)$ has zero bias and, hence,
smaller mean squared error than $\hat\theta$. If, in addition, the
initial estimator $\hat\theta$ has vanishing variance-covariance
matrix as $n\to\infty$ then an application of Chebyschev's inequality
shows that $\tilde\theta$ is consistent, even if $\hat\theta$ is not.
Of course, if $\theta$ is known then there is no reason for
estimation, and furthermore, usually the function $B(\theta)$ cannot
be written in closed-form. The importance of equation~(\ref{estEq}) is
that, despite of its limited practical value, all known methods to
reduce bias can be usefully thought of as attempts to approximate its
solution. These methods can be distinguished into \emph{explicit} and
\emph{implicit}.

\section{Explicit methods}
Explicit methods rely on an \emph{one-step} procedure where
$B(\theta)$ is estimated and then subtracted from $\hat\theta$
resulting in the new estimator $\tilde\theta$. The most popular
explicit methods for reducing bias are the jackknife, the bootstrap
and methods which use approximations of the bias function through
asymptotic expansions of $B(\theta)$.

\subsection{Jackknife}
For many common estimators including the maximum likelihood estimator,
the bias function can be expanded in decreasing powers of
$n$ as
\begin{equation}
\label{jackknifen}
B(\theta) = \frac{b(\theta)}{n} +
\frac{b_2(\theta)}{n^2} + \frac{b_3(\theta)}{n^3} + O(n^{-4}) \, ,
\end{equation}
for an appropriate sequence functions $b(\theta),
b_2(\theta)$, $b_3(\theta)$, $\ldots$, and so on, that are $O(1)$ as $n \to \infty$.W
From~(\ref{jackknifen}), the estimator $\hat\theta_{(-j)}$ which
results from leaving the $j$th random variable out of the original set
of $n$ variables has the same bias expansion as in (\ref{jackknifen})
but with $n$ replaced with $n-1$. In light of this observation,
\citet{quenouille:56} noticed that the estimator
\[
\tilde\theta = n\hat\theta - (n-1)\bar\theta \, ,
\]
where $\bar\theta$ is the average of the $n$ possible leave-one-out
estimators $\hat\theta_{(-1)}, \ldots, \hat\theta_{(-n)}$, has bias
expansion $-b_2(\theta)/n^2 + O(n^{-3})$
which is of smaller asymptotic order than the $O(n^{-1})$ bias of
$\hat\theta$. This procedure is called jackknife \citep[see,][for an
overview of jackknife]{chernick:12}. \citet[\S
2.3]{efron:82} shows the basic geometric argument behind the
jackknife; the jackknife is estimating the bias based on a linear
extrapolation of the expected value of the estimator as a function of
$1/n$.  The same procedure can be carried out for correcting bias in
higher orders essentially replacing the linear extrapolation by
quadratic extrapolation and so on. \citet{schucany:71} give an elegant
way of deriving such higher order corrections in bias with the
jackknife being a prominent special case of their method. The
jackknife is an explicit method because the new estimator
$\tilde\theta$ is simply
\[
\tilde\theta = \hat\theta - B^{(jack)}\, ,
\]
where $B^{(jack)} = (n - 1)(\bar\theta - \hat\theta)$ is the jackknife
estimator of the bias.

\subsection{Bootstrap}
Another class of popular explicit methods for the correction of the
bias comes from the bootstrap framework. Bootstrap is a collection of
methods that can be used to improve the accuracy of inference and
operates under the principle that the ``\emph{bootstrap sample} is for
the sample, what the sample is for the population''. Then the same
procedures that are applied on the sample can equally well be applied
on the bootstrap sample giving direct access to estimated sampling
distributions of statistics \citep[see,][for an overview of
bootstrap]{hesterberg:11}. The two dominant ways to obtain a bootstrap
sample are i) by sampling from the empirical distribution function
(hence sampling with replacement from the original sample) giving rise
to nonparametric bootstrap methods, and ii) by sampling from the
fitted parametric model giving rise to parametric bootstrap
methods. In all cases, the bias of an estimator can be estimated by
$B^{(boot)}= \bar\theta^* - \hat\theta$, where $\bar\theta^*$ is the
average of the estimates based on each of the bootstrap
samples. \citet{efron:93} and \citet{davison:97} are thorough
treatments of bootstrap methodology. Under general conditions
\citet{hall:88} showed that, if $\hat\theta$ has a bias of $O(n^{-1})$
which can be consistently estimated, then the estimator
\begin{equation*}
\tilde\theta = \hat\theta - B^{(boot)} =
2\hat\theta - \bar\theta^*\,.
\end{equation*}
has $O(n^{-2})$ bias.

The estimates of the bias in the gasoline yield data case study were obtained by simulation from the fitted model in Table~\ref{gasolineML}, and thus are parametric bootstrap estimates of the bias.


\subsection{Asymptotic bias correction}
Another widely used class of explicit methods involves the
approximation of $B(\theta)$ by $b(\hat\theta)/n$ which is the
first-term in the right hand side of (\ref{jackknifen}) evaluated at
$\hat\theta$. \citet{cox:68}, in their investigation of higher order
properties of residuals in general parametric models, derive an
expression for the first-order bias term $b(\theta)/n$ in expression
(\ref{jackknifen}), when $\hat\theta$ is the maximum likelihood
estimator. That expression has sparked a still-active research stream
in correcting the bias by using the estimator
\[
\tilde\theta = \hat\theta - b(\hat\theta)/n \,.
\]
\citet{efron:75} showed that $\tilde\theta$ has bias of order
$o(n^{-1})$ which is of smaller order than the $O(n^{-1})$ bias of the
maximum likelihood estimator and that the asymptotic variance of any
estimator with $O(n^{-2})$ bias is greater or equal to the asymptotic
variance of $\tilde\theta$ (second-order efficiency). For the
interested reader, \citet[][\S 9.4]{pace:97} give a thorough
discussion of those properties.

Landmark studies in the literature for asymptotic bias corrections are
\citet{cook:86} who investigate correcting the bias in nonlinear
regression models with Normal errors and \citet{cordeiro:91} who treat
generalized linear models with interesting results on the shrinkage
properties of the reduced-bias estimators in binomial regression
models and an attractive implementation through one supplementary
re-weighted least squares iteration. Furthermore, \citet{botter:98}
and \citet{cordeiro:08} extend the results in \citet{cordeiro:91} and
derive the first-order biases for generalized linear and nonlinear
models with dispersion covariates.

The general form of the first-order bias term of the maximum
likelihood estimator can be found in matrix form in
\citet{kosmidis:10}. Specifically,
\[
\frac{b(\theta)}{n} = - \left\{F(\theta)\right\}^{-1}A(\theta)\, ,
\]
where $A(\theta)$ is a $p$-dimensional vector with components
\begin{equation}
\label{adjustment}
A_t(\theta) = \frac{1}{2}\trace\left[\{F(\theta)\}^{-1} \left\{
    P_t(\theta) + Q_t(\theta) \right\}\right] \quad (t = 1, \ldots,
p) \, ,
\end{equation}
and where
\begin{align*}
  P_t(\theta) & =  E_\theta\{S(\theta)S^\top(\theta)S_t(\theta)\}
  \quad (t = 1, \ldots, p)\, ,  \\
  Q_t(\theta) & = -E_\theta\left\{I(\theta)S_t(\theta) \right\} \quad (t =
  1, \ldots, p)\, ,
\end{align*}
are higher-order joint null moments of the gradient and the matrix of second derivatives of the log-likelihood.

\citet{breslow:95} derive the expressions for the asymptotic biases in
generalized linear mixed models for various estimation methods and
used those to correct for the bias. Higher order corrections have also
appeared in the literature \citep{cordeiro:07} where expressions for
$b(\theta)/n + b_2(\theta)/n^2$ in (\ref{jackknifen}) are
obtained. The expressions involved for such higher-order corrections
are too cumbersome requiring enormous effort in derivation and
implementation, and there is always the danger that the benefits in
estimation from this effort are only marginal, if any, compared to
methods that are based on simply removing the first-order bias term.

\subsection{Advantages and disadvantages of explicit methods}
The main advantage of all explicit methods is the simplicity of their
application. Once an estimate of bias is available, reduction of bias
is simply a matter of an one-step procedure where the estimated bias
is subtracted from the estimates. Nevertheless because of their
explicit dependence on $\hat\theta$, explicit methods directly inherit
any of the instabilities of the original estimator.
Such cases involve models with categorical responses where there is a
positive probability that the maximum likelihood estimator is not
finite \citep[see,][for conditions that characterize when infinite
estimates occur in multinomial response models]{albert:84} and have
been the subject of study in works like \citet{mehrabi:95},
\citet{heinze:02}, \citet{bull:02}, \citet{kosmidis:09},
\citet{kosmidis:11} and \citet{kosmidis:13}. In particular,
\citet{kosmidis:13} relates to the case study of the proportional odds
models, discussed below.

Furthermore, asymptotic bias correction methods have the disadvantage
that are only applicable when $b(\theta)/n$ can be obtained in
closed-form, which can be a tedious or even impractical task for many
models \citep[see, for example,][where the expressions for
$b(\theta)/n$ are given for Beta regression models]{grun:12}.

\section{Implicit methods}
Implicit methods approximate $B(\theta)$ at the target estimator
$\tilde\theta$ and then solve equation~(\ref{estEq}) with respect to
$\tilde\theta$. Hence, $\tilde\theta$ is the solution of an implicit
equation.

\subsection{Indirect inference}
Indirect inference is a class of inferential procedures that appeared
in the Econometrics literature in \citet{gourieroux:93} and can be
used for bias reduction. The simplest approach to bias reduction via
indirect inference attempts to solve the equation
\[
\tilde\theta = \hat\theta - B(\tilde\theta) \, ,
\]
by approximating $B(\theta)$ at $\tilde\theta$ through parametric
bootstrap. \citet{kuk:95} independently produced the same idea for
reducing the bias in the estimation of generalized linear models with
random effects and \citet{jiang:04} give a comprehensive review of
indirect inference from a statistical point of view. Furthermore,
\citet{gourieroux:00} and \citet{phillips:12} discuss bias reduction
through indirect inference in econometric applications.
\citet{pfeffermann:12} give an alternative approach to bias
reduction which is in line with the basic idea of indirect inference.

\subsection{Bias-reducing adjusted score equations}
For the case where $\hat\theta$ is the maximum likelihood estimator
and under the usual regularity conditions, \citet{firth:93} and
\citet{kosmidis:09} investigate what penalties need to be added to
$S(\theta)$ in order to get an estimator that has asymptotically
smaller bias than that of the maximum likelihood estimator. In its
simplest form such an approach requires finding $\tilde\theta$ by
solving the adjusted score equations
\begin{equation}
\label{firth}
S(\tilde\theta) + A(\tilde\theta) = 0\, ,
\end{equation}
with $A(\theta)$ as given in (\ref{adjustment}).
Then $\tilde\theta$ is an estimator with $o(n^{-1})$
bias. Equation~(\ref{firth}) can be rewritten as
\[
\{F(\tilde\theta)\}^{-1}S(\tilde\theta)  = \frac{b(\tilde\theta)}{n}
\, ,
\]
which reveals that $\tilde\theta$ is another approximate solution to
equation (\ref{estEq}) because $b(\tilde\theta)/n$ approximates
$B(\tilde\theta)$ up to order $O(n^{-2})$ and
$\{F(\theta)\}^{-1}S(\theta)$ is the $O(n^{-1/2})$ term in the
asymptotic expansion of $\hat\theta - \theta$ evaluated at
$\theta:=\tilde\theta$.

An important property of the estimator based on adjusted score
functions, which is also shared by the estimator from asymptotic bias
correction, is that it has the same asymptotic distribution as the
maximum likelihood estimator, namely a Normal distribution centered at
the true parameter value with variance-covariance matrix the
Cram\'{e}r-Rao lower bound $\{F(\theta)\}^{-1}$. Hence, the
first-order methods that are used for the maximum likelihood
estimator, like Wald-type confidence intervals, score tests for model
comparison, and so on, are unaltered in their form and apply directly
by using the new estimators.

It is noteworthy that in the case of full exponential families (for
example, logistic regression and Poisson log-linear models) the
solution of (\ref{firth}) can be obtained by direct maximization of a
penalized likelihood where the penalty is the Jeffreys
\citep{jeffreys:46} invariant prior \citep[see,][for
details]{firth:93}. It should also be stressed that not all models
admit a penalized likelihood interpretation of bias reduction via
adjusted scores. \citet{kosmidis:09} give an easy-to-check necessary
and sufficient condition that identifies which univariate generalized
linear models admit such penalized likelihood interpretation and
provide the form of the resultant penalties when the condition
holds. That condition is a restriction on the variance function of the
responses in terms of the derivative of the chosen link
function.

\subsection{Advantages and disadvantages}
The main disadvantage of implicit methods is that their application
requires the solution of a set of implicit equations which in most of
the useful cases requires numerical optimization. This task is even
more computationally demanding for indirect inference approaches in
general models because of the necessity to approximate the bias
function in a $p$ dimensional space. Furthermore, indirect inference
approaches inherit the disadvantages of explicit methods because they
explicitly depend on the original estimator.

The approach in \citet{firth:93} and \citet{kosmidis:09}, on the other
hand, does not directly depend on $\hat\theta$ and hence has gained
considerable attention compared to the other approaches. Another
reason for the considerable adaptation of this method are recent
advances which simplify application through either iterated
first-order bias adjustments (see, \citealt{grun:12, kosmidis:10}) or
iterated maximum likelihood fits on pseudo observations
\citep[see,][]{kosmidis:09, kosmidis:11, kosmidis:13}. Of course, as
for the asymptotic bias correction methods the adjusted score equation
approach to bias reduction has the disadvantage of being directly
applicable only under the same conditions that guarantee the good
limiting behaviour of the maximum likelihood estimator and only when
the score functions, Fisher information and the first-order bias term
of the maximum likelihood estimator are available in closed form.

\section{Proportional odds models}
This example was analysed in \citet[][]{kosmidis:13}. The data set in
Table~\ref{wineData} is from \citet{randall:89} and concerns a
factorial experiment for investigating factors that affect the
bitterness of white wine. There are two factors in the experiment,
temperature at the time of crushing the grapes (with two levels,
``cold'' and ``warm'') and contact of the juice with the skin (with
two levels ``Yes'' and ``No''). For each combination of factors two
bottles were rated on their bitterness by a panel of 9 judges. The
responses of the judges on the bitterness of the wine were taken on a
continuous scale in the interval from 0 (``None'') to 100
(``Intense'') and then they were grouped correspondingly into $5$
ordered categories, $1$, $2$, $3$, $4$, $5$.

\begin{table}[t!]
  \caption{The wine tasting data \citep{randall:89}.}
  \begin{center}
    \begin{tabular}{ccccccc}
      \toprule
      Temperature & Contact &
      \multicolumn{5}{c}{Bitterness scale} \\
      & & 1 & 2 & 3 & 4 & 5 \\ \midrule
      Cold & No  & 4 & 9 & 5 & 0 & 0 \\
      Cold & Yes & 1 & 7 & 8 & 2 & 0 \\
      Warm & No  & 0 & 5 & 8 & 3 & 2 \\
      Warm & Yes & 0 & 1 & 5 & 7 & 5 \\ \bottomrule
    \end{tabular}
    \end{center}
    \label{wineData}
  \end{table}

\begin{table}[t]
  \caption{The maximum likelihood and the reduced-bias estimates for the parameters of
    model~(\ref{partialProp}), the corresponding estimated standard
    errors (in parenthesis) and the values of the corresponding $Z$ statistic
    for the hypothesis that the corresponding parameter is
    zero. The maximum likelihood estimates and Z-statistics are as reported by the \texttt{clm} R package \texttt{ordinal} \citep{ordinal:12}.}
  \begin{center}
  \begin{tabular}{crrrrrr}
      \toprule
      & \multicolumn{3}{c}{Maximum likelihood} &
      \multicolumn{3}{c}{Adjusted score equations} \\
      Parameter & \multicolumn{2}{c}{Estimates} & $Z$-statistic & \multicolumn{2}{c}{Estimates} & $Z$-statistic \\ \midrule
      $\alpha_1$ & -1.27 & (0.51) & -2.46 & -1.19 & (0.50) & -2.40 \\
      $\alpha_2$ &  1.10 & (0.44) & 2.52 &  1.06 & (0.44) & 2.42 \\
      $\alpha_3$ &  3.77 & (0.80) & 4.68 &  3.50 & (0.74) & 4.73 \\
      $\alpha_4$ & 28.90 & (193125.63) & 0.00 &  5.20 & (1.47) & 3.52 \\
      $\beta_1$ & 25.10 & (112072.69) & 0.00 &  2.62 & (1.52) & 1.72 \\
      $\beta_2$ &  2.15 & (0.59) & 3.65 &  2.05 & (0.58) & 3.54 \\
      $\beta_3$ &  2.87 & (0.82) & 3.52 &  2.65 & (0.75) & 3.51 \\
      $\beta_4$ & 26.55 & (193125.63) & 0.00 &  2.96 & (1.50) & 1.98 \\
      $\theta$ & 1.47 & (0.47) & 3.13 & 1.40 & (0.46) & 3.02 \\
      \bottomrule
    \end{tabular}
    \end{center}
    \label{wineDataEsts}
\end{table}

The task of the analysis is to check whether there are departures from
the assumptions of the proportional odds models that is well-used in
the analysis of ordinal responses. For performing such a test we use
the more general partial proportional odds model of
\citet{peterson:90} with
\begin{equation}
  \label{partialProp}
  \log{\frac{\gamma_{rs}}{1 - \gamma_{rs}}} = \alpha_s - \beta_s w_r -
  \theta z_r \quad (r = 1, \ldots, 4; s = 1, \ldots, 4)\, ,
\end{equation}
where $w_r$ and $z_r$ are dummy variables representing the factors
temperature and contact, respectively, $\alpha_1, \ldots, \alpha_4,
\beta_1, \beta_2, \beta_3, \beta_4, \theta$ are model parameters and $\gamma_{rs}$ is the
cumulative probability for the $s$th category at the $r$th combination
of levels for temperature and contact. Then we can check for departures
from the proportional odds assumption by testing the hypothesis
$\beta_1 = \beta_2 = \beta_3 = \beta_4$, effectively comparing
(\ref{partialProp}) to the proportional odds nested model that is
implied by the hypothesis.

Table~\ref{wineDataEsts} shows the maximum likelihood estimates for
model~(\ref{partialProp}) and the corresponding estimated standard
errors as reported by the \texttt{clm} function of the R package
\texttt{ordinal} \citep{ordinal:12}. It is directly apparent that the
absolute value of the estimates and estimated standard errors for the
parameters $\alpha_4$, $\beta_1$ and $\beta_4$ is very
large. Actually, these would diverge to infinity as the stopping
criteria of the iterative fitting procedure used become stricter and
the number of allowed iterations increases. The estimates for the
remaining parameters are all finite and will preserve the value shown
in Table~\ref{wineDataEsts} even if the number of allowed iterations
increases. This is an instance of the problems that practitioners may
face when dealing with categorical response models. Using a Wald-type
statistic based on the maximum likelihood estimator for testing the
hypothesis of proportional odds would be adventurous here because such
a statistic explicitly depends on the estimates of $\beta_1$,
$\beta_2$, $\beta_3$ and $\beta_4$. Of course, given that the
likelihood is close to its maximal value at the estimates in
Table~\ref{wineData}, a likelihood ratio test can be used instead; the
likelihood ratio test for this particular example has been carried out
in \citet[][\S 7]{christensen:12}.

Note here that methods like the bootstrap and jackknife would require
special considerations for their application in a well-designed
experiment like the above, the question to be answered being what
consists an observation to be re-sampled or left-out. Even if such
considerations were resolved, bootstrap and jackknife would be prone
to the problem of infinite estimates. The latter is also true for the
estimator based on asymptotic bias corrections and for indirect
inference.

\citet{kosmidis:13} derives the adjusted score equations for
cumulative link models, and uses them to calculate the reduced-bias
estimates shown in the right of Table~(\ref{wineDataEsts}). The
reduced-bias estimates based on the adjusted score functions are
finite and, through the asymptotic normality of the reduced-bias
estimator, they can form the basis of a Wald-test for the hypothesis
$\beta_1 = \beta_2 = \beta_3 = \beta_4$. This test has been carried
out in \citet{kosmidis:13} and gives a $p$-value of $0.861$, providing
no evidence against the hypothesis of proportional odds.

Furthermore, the values of the $Z$-statistics for $\alpha_4$,
$\beta_1$ and $\beta_4$ in Table~\ref{wineDataEsts} are essentially
zero when based on the maximum likelihood estimator. This is typical
behaviour when the estimates diverge to infinity and it happens
because the estimated standard errors diverge much faster than the
estimates, irrespective of whether or not there is evidence against
the individual hypotheses. This is also true if we were testing
individual hypothesis at values other than zero, and can lead to
invalid conclusions if the maximum likelihood output is interpreted
naively; as shown in Table~\ref{wineDataEsts}, the Z-statistics
based on the reduced-bias estimates are far from being zero.

Such inferential pitfalls with the use of the maximum likelihood
estimator are not specific to partial proportional odds models. For
most models for categorical and discrete data (binomial response
models like the logistic regression, multinomial response models,
Poisson log-linear models, and so on) there is a positive probability
of infinite estimates. Bias reduction through adjusted score functions
has been found to provide a solution to those problems and the
corresponding methodology is quickly gaining in popularity and has
found its way to commercial software like Stata and SAS. Open-source
solutions include the \texttt{logistf} R package \citep{logistf:13}
for logistic regressions which is based on the work in
\citet{heinze:02}, the \texttt{pmlr} R package \citep{pmlr:10} for
multinomial logistic regressions based on the work of \citet{bull:02},
and the \texttt{brglm} R package \citep{brglm:13, kosmidis:09a} which
at the time of writing handles all binomial-response models. At the
time of writing, the \texttt{brglm} R package is being extended for
the next major update which will handle all generalized linear models,
including multinomial logistic regression \citep{kosmidis:11} and
ordinal response models \citep{kosmidis:13}.

\section{Gasoline yield data revisited}
In this section, the reduced-bias estimates for the parameters of
model~(\ref{gasolineModel}) are calculated using jackknife, bootstrap,
asymptotic bias correction and the approach of bias-reducing adjusted
score functions. The full parametric bootstrap estimate of the bias
has been obtained in our earlier treatment showing that the bias on
the regression parameters is of no consequence. A fully non-parametric
bootstrap where the bootstrap samples are produced by sub-sampling
with replacement the full response-covariate combinations $(y_i,
s_{i1}, \ldots, s_{i9}, t_i)$ $(i = 1, \ldots, n)$ is not advisable
here because the 9 dummy variables $s_{i1}, \ldots, s_{i9}$ are
representing $10$ distinct experimental settings and sub-sampling
those will result in singular fits with high probability \citep[see,
also][\S 6.3 for a description of such problems in the simpler case of
multiple linear regression]{davison:97}. An intermediate sub-sampling
strategy is to resample residuals and use them with the original model
matrix to get samples for the response. This strategy lies between
fully non-parametric bootstrap and fully parametric bootstrap
\citep[see,][\S 7]{davison:97}.  Residual re-sampling works well in
multiple linear regression because the response is related linearly to
the regression parameters, which is not true for Beta
regression. For more complicated models like generalized linear models
and Beta regression, an appropriate residual definition has to be
chosen. Because Beta responses are restricted in $(0,1)$, the best
option is to resample residuals on the scale of the linear predictor
and then transform back to the response scale using the inverse of the
logistic link, obtaining bootstrap samples for the response
\citep[see,][expression (7.13) for rationale and
implementation]{davison:97}. In the current case we choose the
``standardized weighted residual 2'' of \citet{espinheira:08} because
it appears to be the one that is least sensitive to the inherent
skewness of the response.

The reduced-bias estimates of $\phi$ using jackknife,
residual-resampling bootstrap (with 9999 bootstrap samples),
asymptotic bias correction and bias-reducing adjusted score functions
are $165.682$, $236.003$, $261.206$ and $261.038$, respectively, all
indicating that the maximum likelihood estimator of $\phi$ is prone to
substantial upward bias.  The simulations in \citet{kosmidis:10}
illustrate that asymptotic bias correction and the bias-reducing
adjusted score functions, correctly inflate the estimated standard
errors to the extent that almost the exact coverage of the first-order
Wald-type confidence intervals is recovered.

\section{Discussion and conclusion}
As can be seen from the earlier case-studies, reduced-bias estimators
can form the basis of asymptotic inferential procedures that have
better performance than the corresponding procedures based on the
initial estimator. \citet{heinze:02}, \citet{bull:07},
\citet{kosmidis:07}, \citet{kosmidis:09, kosmidis:10}, and
\citet{grun:12} all demonstrate that such improved procedures are
delivered either by using the penalized likelihood that results from
the approximation of equation~(\ref{estEq}), or by replacing the
initial estimator with the reduced-bias estimator in Wald-type pivots,
as was done in the case studies of this review.

At the time of writing the current review there is no general answer
to which of the methods that have been reviewed here produces better
results. All methods deliver estimators that have $o(n^{-1})$ bias
which is asymptotically smaller than the $O(n^{-1})$ bias of the
maximum likelihood estimator. In models with categorical or discrete
responses, the adjusted score equations approach is preferable to the
other bias reduction approaches because the resultant estimates appear
to be always finite, even in cases where the maximum likelihood
estimates are infinite \citep[see, for example][for generalized linear
and non-linear models with binomial, multinomial and Poisson
responses]{heinze:02, bull:07, kosmidis:09, kosmidis:11,
  kosmidis:13}. This has led researchers to promoting the routine use
of the adjusted score equations in such models as an improved
alternative to maximum likelihood.

The general use, though, of the adjusted score equations approach is
limited by its dependence on a closed-form expression for the
first-order bias of the maximum likelihood estimator which may not be
readily available or even intractable (for example, generalized linear
mixed effects models).

At this point, we should also stress that improving bias does not
always have desirable effects; an improvement in bias can sometimes
result in inflation of the mean squared error, through an inflation in
the estimator's variance. The use of simple simulation studies,
similar to the one in \citet{kosmidis:10} is recommended for checking
whether that is the case. If that is the case then the use of
reduced-bias estimates in test statistics and confidence intervals is
not recommended.

Furthermore, bias is a parameterization-specific quantity and any
attempt to improve it will violate the invariance properties of the
maximum likelihood estimator. Hence, bias-reduction methods should be
used with care, unless the parameterization is fixed either by the
context or by the practitioner.

All the discussion in the current review has focused on the effect
that bias can have and the benefits of its reduction in cases where
the usual regularity conditions are satisfied. An important research
avenue is the reduction of bias under departures from the regularity
conditions and especially when the dimension of the parameter space
increases with the sample size.  \citet{lancaster:00} gives a review
of the issues that econometricians and applied statisticians face in
such settings. A viable route towards reduction of bias in such cases
comes from the use of modified profile likelihood methods
\citep[see,][for a brief introduction]{reid:10}, which have been
successfully used for reducing the bias in the estimation of dynamic
panel data models in \citet{bartolucci:12}. Another route is the
appropriate adaptation of indirect inference approaches or of other
approximate solutions of equation~(\ref{estEq}) in such settings. The
Econometric community is currently active in this direction, with a
recent example being \citet{gourieroux:10} where indirect inference is
applied to dynamic panel data models. These early attempts are only
indicative that, there is still much to be explored and much work to
be done on the topic of bias reduction in parametric estimation.

\section{Acknowledgement}
The author thanks the Editor, the Review Editor and three anonymous
Referees for detailed, helpful comments which have substantially
improved the presentation.

\section{Supplementary material}
Supplementary material for this article is available upon request and
includes a script that reproduces the analyses of the gasoline yield
data using Beta regression models ({\tt wireGasoline.R}). The script
for the wine tasting data case study with cumulative link models is
available at the supplementary material of \citet{kosmidis:13} and can
be obtained following the associated DOI link in the References
section. A change in behaviour at the latest version of the {\tt
  ordinal} R package at the time of writing (version 2013.9-30)
returns non-available values for the estimates, estimated standard
errors and Z-values for the parameters of cumulative link models when
at least one infinite estimate is detected. For this reason, the wine
data analyses can be reproduced using version {\tt 2012.01-19} of the
{\tt ordinal} package which is available for download at {\tt
  http://cran.r-project.org/src/contrib/Archive/ordinal/}.

\bibliographystyle{chicago}

\end{document}